\documentclass[conference]{IEEEtran}
\IEEEoverridecommandlockouts
\usepackage{cite}
\usepackage{amsmath,amssymb,amsfonts}
\usepackage{algorithmic}
\usepackage{graphicx}
\usepackage{textcomp}
\usepackage{xcolor}
\def\BibTeX{{\rm B\kern-.05em{\sc i\kern-.025em b}\kern-.08em
    T\kern-.1667em\lower.7ex\hbox{E}\kern-.125emX}}

\usepackage[T1]{fontenc} % add special characters (e.g., umlaute)
\usepackage[utf8]{inputenc} % set utf-8 as default input encoding
\usepackage{color}
\usepackage{acronym}
\usepackage{booktabs}
\usepackage{amssymb}
\usepackage{pifont}
\usepackage{multirow}
\usepackage{siunitx}
\usepackage{url}
\usepackage{hyperref}
\hypersetup{
    colorlinks=true,
    urlcolor=blue
}

\newcommand{\LL}{\mathcal{L}}

\newcommand{\RR}{\mathbb{R}}

\newcommand{\cc}{\mathbf{c}}
\newcommand{\h}{\mathbf{h}}
\newcommand{\x}{\mathbf{x}}
\newcommand{\xn}{\mathbf{x}^{(n)}}

\newcommand{\qn}{\mathbf{q}^{(n)}}
\newcommand{\z}{\mathbf{z}}
\newcommand{\zn}{\mathbf{z}^{(n)}}
\newcommand{\xbar}{\mathbf{\bar{x}}}
\newcommand{\zbar}{\mathbf{\bar{z}}}
\newcommand{\ztilde}{\mathbf{\tilde{z}}}
\newcommand{\s}{\mathbf{s}}
\newcommand{\sbar}{\mathbf{\bar{s}}}

\acrodef{EMA}{Exponential Moving Average}
\acrodef{JEPA}{Joint-Embedding Predictive Architectures}
\acrodef{ViT}{Vision Transformer}
\acrodef{SSL}{self-supervised learning}
\acrodef{MIR}{Music Information Retrieval}
\acrodef{MLP}{Multi-Layer Perceptron}
\acrodef{LMS}{Log-scaled Mel-Spectrograms}

\title{Zero-shot Musical Stem Retrieval\\with Joint-Embedding Predictive Architectures}

\author{
    \IEEEauthorblockN{
        Alain Riou$^{1,2}$, Antonin Gagneré$^1$, Gaëtan Hadjeres$^3$, Stefan Lattner$^2$, Geoffroy Peeters$^1$
    }
    \IEEEauthorblockA{
        \textit{$^1$LTCI, Télécom-Paris, Institut Polytechnique de Paris, France} \\
        \textit{$^2$Sony Computer Science Laboratories - Paris, France} \\
        \textit{$^3$Sony AI, Zurich, Switzerland} \\
        alain.riou@sony.com
    }
}

\begin{document}
%\ninept
%
\maketitle

\begin{abstract}
    % Musical stem compatibility estimation is interesting for mashup creation, loop/stem retrieval and musical accompaniment generation.
    % In this paper, we introduce a new method based on JEPAs, where an encoder and a predictor are jointly trained to produce latent representations of a context and predict latent representations of a target.
    % In particular, we design our predictor to be able 
    % Finally, we show that pretraining the encoder with contrastive learning significantly improves the quality of the model.
    
    % We evaluate our model on a variety of retrieval tasks. \TODO{write abstract}

    In this paper, we tackle the task of musical stem retrieval.
    Given a musical mix, it consists in retrieving a stem that would fit with it, i.e., that would sound pleasant if played together.
    To do so, we introduce a new method based on Joint-Embedding Predictive Architectures, where an encoder and a predictor are jointly trained to produce latent representations of a context and predict latent representations of a target.
    In particular, we design our predictor to be conditioned on arbitrary instruments, enabling our model to perform zero-shot stem retrieval.
    In addition, we discover that pretraining the encoder using contrastive learning drastically improves the model's performance.
    
    We validate the retrieval performances of our model using the MUSDB18 and MoisesDB datasets.
    We show that it significantly outperforms previous baselines on both datasets, showcasing its ability to support more or less precise (and possibly unseen) conditioning.
    We also evaluate the learned embeddings on a beat tracking task, demonstrating that they retain temporal structure and local information.
\end{abstract}

{\normalsize
\textbf{Disclaimer:}
\textit{This paper is an extended version of our ICASSP 2025 publication. The ICASSP version is available at \texttt{\url{https://arxiv.org/pdf/2411.19806v1}}.}
}

\section{Introduction}

Musical stem retrieval consists of finding a stem that fits when played together with a reference audio mix.
It relies on finding a good way to measure the compatibility between different audio samples, considering elements such as tempo, tonality, and various stylistic features.

Early research on this topic focuses on automatic mashup creation, retrieving full songs rather than individual stems.
It computes a ``mashability'' score using algorithms for beat tracking and chord estimation, then selects the song that maximizes this score~\cite{AutoMashupper}. However, these methods only consider rhythmic and harmonic features, overlooking timbre and playing style, which are crucial aspects of musical compatibility.
Moreover, these approaches are sensitive to errors made by the underlying algorithms and require full mixes (chord estimation on a drum stem returns unreliable results), making them unsuitable for single-stem retrieval~\cite{StemJEPA}.

Later, Siamese networks~\cite{Siamese} and \ac{SSL} emerged as a promising way to model the similarity between data samples without relying on handcrafted rules.
Given a neural network that projects input data in a metric latent space, the core idea is to present it with pairs of related inputs and train it to map each pair to nearby points within this space, usually by optimizing a contrastive loss~\cite{SimCLR}.
This approach has shown success in producing informative representations in various domains such as image~\cite{SimCLR}, audio~\cite{CLMR,COLA}, and multimodal data~\cite{CLIP,CLAP}.
In particular, it has also been applied to compatibility estimation between loops~\cite{Chen2020}, and to drum sample retrieval by using tracks and their drum samples as positive pairs~\cite{SampleMatch}.

%\TODO{say that it's cool to have conditioning}
%However, in music production, a user may not want to retrieve \emph{any} stem that fits a given mix but also to request specific instruments or aesthetics.
More recently, Stem-JEPA~\cite{StemJEPA} proposes using \ac{JEPA} for stem retrieval.
%, enabling the use of conditioning to retrieve instruments specifically.
\ac{JEPA}s, indeed, involve training two networks—a context encoder and a predictor—to jointly predict the target’s representation from the context's one.
A key feature of these architectures is that the predictor can support additional conditioning, enabling the learning of richer latent spaces than with previous contrastive methods~\cite{BYOL,IJEPA,M2D}.
In particular, \cite{StemJEPA} performs stem retrieval by conditioning the predictor on the instrument class of the target stem. However, their model is limited to four instruments, restricting its practical usability.

In this paper, we build upon these advancements to propose a new JEPA-based model for musical stem retrieval. Our model is more flexible, handling any music input and allowing conditioning on any instrument provided as free-form text. Additionally, we incorporate a contrastive pretraining phase before training the full \ac{JEPA} itself and show that it significantly boosts performance. We also replace the conditioning method from \cite{StemJEPA} with FiLM-inspired conditioning~\cite{FiLM}.

We evaluate our model on retrieval tasks using two source separation datasets: MUSDB18, whose tracks are divided into four standard stems~\cite{MUSDB18}, and MoisesDB, which contains more stems and for which annotations are hierarchical and more precise~\cite{MoisesDB}. This allows us to validate the effectiveness of our design choices across different levels of conditioning granularity.
Finally, we evaluate the embeddings learned by our model on a beat tracking task, highlighting that they retain some temporal information, making them potentially usable for tasks beyond stem retrieval.
We make our code publicly available.\footnote{\texttt{\url{https://github.com/SonyCSLParis/Stem-JEPA}}}

\section{Method}

\begin{figure*}
    %\centering
    % \captionsetup{font=\fontsize{4}{11}\selectfont}
    \includegraphics[width=1.005\textwidth]{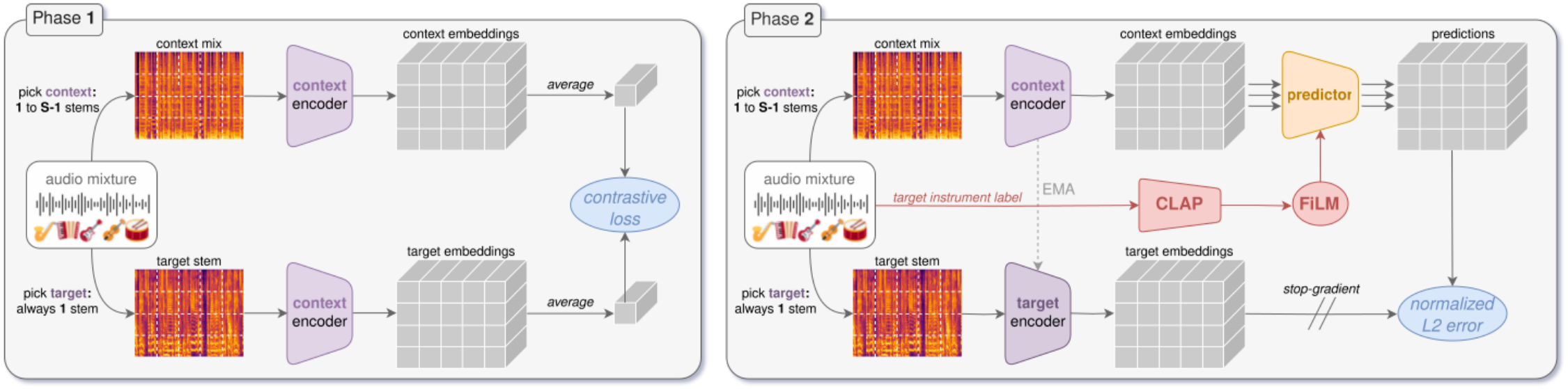}
    \caption{
        Overview of our model.
        In both phases, it is trained using a pair composed of a context mix $\x$ and a target stem $\xbar$ extracted from the same track.
        In Phase 1 (left), we pretrain only the encoder $f_{\theta}$ using contrastive learning to bring together the (averaged) representations $\s$ and $\sbar$ of the context and target.
        In Phase 2 (right), a predictor $g_{\phi}$, conditioned on the instrument of the target stem $c$, tries to retrieve the patchwise representations of the target stem $\zbar$ from the ones of the context mix $\z$. In this phase, the parameters of the target encoder $f_{\bar{\theta}}$ are updated as an \ac{EMA} of the ones of the encoder $f_{\theta}$.}
    \label{fig:model}
\end{figure*}

Our model comprises two trainable networks: an encoder $f_{\theta}$ with parameters $\theta$ and a predictor $g_{\phi}$ with parameters $\phi$.
The encoder is a \ac{ViT} \cite{ViT} that takes \ac{LMS} as input and returns a grid of latent embeddings, while the predictor is a \ac{MLP} that acts separately on each of these embeddings.
These two networks are trained in two successive phases, as depicted in Figure \ref{fig:model}.

\subsection{Phase 1: Contrastive pretraining}

First, we pretrain only the encoder $f_{\theta}$ using contrastive learning.
Given a chunk of audio composed of $S$ stems, we first pick one as the \emph{target} $\xbar$ and a random subset of the remaining ones as the \emph{context} $\x$.
$\x$ and $\xbar$ are then turned into \ac{LMS}, divided into a regular time and frequency grid of $K$ non-overlapping patches, and fed into the encoder $f_{\theta}$.
This encoder returns sequences of patch-wise embeddings $\z, \zbar \in \RR^{K \times d}$, with $d$ being the latent dimension.

We then use these embeddings' mean over $K$ as the latent representations $\s, \sbar \in \RR^d$ of the context mix and target stem, respectively.

Finally, the parameters $\theta$ of the encoder are optimized by minimizing the contrastive loss $\LL_c$ within batches of representations, using $(\s, \sbar)$ as a positive pair and the other elements of the batch $B$ as negative samples, as in \cite{SimCLR}:
\begin{equation}
    \LL_c(\s, \sbar) = - \log \left(\dfrac{\exp (\text{sim} (\s, \sbar) / \tau)}{\sum_{\s' \in B \setminus \{\s\}} \exp ( \text{sim} (\s, \s') / \tau)} \right),
\end{equation}
where $\text{sim}$ denotes cosine similarity and $\tau$ is a trainable parameter.

By pushing together mixes and stems from the same song, our model thus learns a latent space where nearby samples are likely to be musically compatible.
%One can retrieve stems via a nearest-neighbor search in that space, as in \cite{SampleMatch}.
%However, it does not able

\subsection{Phase 2: Joint-Embedding Predictive Architecture}

The second phase is similar to \cite{StemJEPA}.
We randomly pick a context mix $\x$, a target stem $\xbar$, and turn them into \ac{LMS}. 
We denote the instrument label of the target stem as $c$.
The context mix passes through the encoder $f_{\theta}$, which returns a sequence of embeddings $\z = (\z_1, \dots, \z_K) \in \RR^{K \times d}$.
Conversely, we obtain $\zbar = (\zbar_1, \dots, \zbar_K) \in \RR^{K \times d}$ by passing the target stem $\xbar$ through a \emph{target encoder} $f_{\bar{\theta}}$ with parameters $\bar{\theta}$.%\footnote{Both encoders are initialized with the weights of the trained encoder from phase 1.}

Then, a predictor $g_{\phi}$, conditioned on the target instrument label $c$, produces predictions $\ztilde_i \in \RR^d$ from the context embeddings, i.e., for all $k \in \{ 1, \dots, K \}$:
%\TODO{Stefan: shouldn't $\ell$ be also a parameter to $g_{\phi}$?}:
\begin{equation}
    \ztilde_k = \ g_{\phi}(\z_k, c).
\end{equation}

The parameters $(\theta, \phi)$ are optimized by minimizing the mean squared error $\LL$ between the (normalized) predictions and target embeddings:
\begin{equation}
    \LL(\ztilde, \zbar) = \sum_{k=1}^{K} \left\| \dfrac{\ztilde_k}{\| \ztilde_k \|} - \dfrac{\zbar_k}{\| \zbar_k \|} \right\|^2.
\end{equation}

Furthermore, the parameters $\bar{\theta}$ of the target encoder are optimized as an \ac{EMA} of $\theta$, i.e.
\begin{equation}
    \bar{\theta}_i = \tau_i \bar{\theta}_{i-1} + (1 - \tau_i) \theta_i,
\end{equation}
where the \ac{EMA} rate $\tau_i$ is linearly interpolated between $\tau_0$ and $\tau_T$, $T$ being the total number of training steps, as in \cite{StemJEPA}.

\subsection{Predictor conditioning}

In \cite{StemJEPA}, it is proposed to condition the predictor by learning a specific embedding for each instrument class.
In \cite{StemJEPA}, the four classes are fixed and correspond to ``bass'', ``drums'', ``vocals'', and ``other''.
This is a major limitation of the model since updating the classes would require retraining the whole system.

We propose replacing this with conditioning on the text embeddings of a pretrained CLAP model~\cite{LAIONCLAP}.
%Our predictor can, therefore, be conditioned on various instrument classes provided as free-form text descriptions ...}
%The main improvement of our model compared to \cite{StemJEPA} lies in the conditioning of the predictor.
%In fact, \cite{StemJEPA} learns a specific embedding for each instrument class, resulting in a fixed set of only four sources (``bass'', ``drums'', ``vocals'', and ``other'').
%
%In contrast, we use the text embeddings of a pretrained CLAP model~\cite{LAIONCLAP}. 
Our predictor can, therefore, be conditioned on various instrument classes provided as free-form text descriptions. Moreover, since CLAP is trained on text/audio pairs, the embeddings of instruments with similar timbre are expected to be close. This is particularly helpful for zero-shot retrieval, i.e., retrieving a stem whose instrument is not in the training set.

Let $\cc = \text{CLAP}(c) \in \RR^p$ be the embedding used for conditioning the predictor.
While in \cite{StemJEPA}, this instrument class embedding is concatenated to the context embeddings $\z_i$, here we instead use FiLM conditioning~\cite{FiLM}.
More precisely, our predictor $g_{\phi}$ is a \ac{MLP} with $L$ layers and $m$ hidden units. For $l \in \{1, \dots, L-1 \}$, we learn two affine mappings $\beta_l$ and $\gamma_l : \RR^p \to \RR^m$. Let $\h_l$ be the output of the $l$-th layer of $g_{\phi}$, the input of the $l+1$-th layer is:
\begin{equation}
    \widehat{\h}_{l+1} = \text{ReLU} \left( \gamma_l(\cc) \cdot \h_l + \beta_l(\cc) \right),
\end{equation}
where $\cdot$ denotes the Hadamard product.

\begin{table*}
    \caption{
        Retrieval performances of our model on the MUSDB18 and MoisesDB datasets.
        For MoisesDB, we report the results both when conditioning the predictor on the precise instrument class of the target stem (fine conditioning) or on the broader corresponding category (coarse conditioning).
        All metrics are in percentages.}
    \label{tab:retrieval}
    	
	%\makebox[0pt][l]{\hspace{-1.6cm}
    {\scriptsize
        \begin{tabular}{lccccccccccccccccccc}
            \toprule
            & \multicolumn{5}{c}{MUSDB18} & & \multicolumn{5}{c}{MoisesDB (fine conditioning)} & & \multicolumn{5}{c}{MoisesDB (coarse conditioning)} \\
            \cmidrule{2-6} \cmidrule{8-12} \cmidrule{14-18}
            & \multicolumn{3}{c}{Recall $\uparrow$} & \multicolumn{2}{c}{Norm. Rank $\downarrow$} & & \multicolumn{3}{c}{Recall $\uparrow$} & \multicolumn{2}{c}{Norm. Rank $\downarrow$} & & \multicolumn{3}{c}{Recall $\uparrow$} & \multicolumn{2}{c}{Norm. Rank $\downarrow$} \\
            Model & R@1 & R@5 & R@10 & mean & median & & R@1 & R@5 & R@10 & mean & median & & R@1 & R@5 & R@10 & mean & median \\
            \midrule
            Contrastive baseline &
            2.2 & 72.5 & 86.0 & 1.5 & 0.7 & &
            2.7 & 31.2 & 43.3 & 7.7 & 0.6 & &
            2.7 & 31.2 & 43.3 & 7.7 & 0.6 \\
            Stem-JEPA~\cite{StemJEPA} &
            33.0 & 63.2 & 76.2 & 2.0 & 0.5 & &
            9.9 & 24.1 & 31.6 & 11.7 & 1.7 & &
            9.9 & 24.1 & 31.6 & 11.7 & 1.7 \\
            \midrule
            +CLAP conditioning &
            33.5 & 63.0 & 74.5 & 2.5 & 0.5 & &
            19.3 & 35.8 & 44.0 & 7.2 & 0.7 & &
            18.9 & 37.8 & 46.9 & 7.3 & 0.5 \\
            +contrastive pretraining &
            32.2 & \textbf{91.0} & \textbf{96.2} & \textbf{0.6} & \textbf{0.3} & &
            12.5 & 45.6 & \textbf{59.3} & \textbf{1.9} & 0.3 & &
            11.1 & 44.5 & 58.1 & \textbf{2.1} & 0.3 \\
            +FiLM conditioning &
            \textbf{38.8} & 89.7 & 95.0 & 0.7 & \textbf{0.3} & &
            \textbf{22.0} & \textbf{49.5} & 57.8 & 4.5 & \textbf{0.2} & &
            \textbf{19.2} & \textbf{49.5} & \textbf{60.0} & 3.8 & \textbf{0.2} \\
            \bottomrule
        \end{tabular}
    }
\end{table*}

\subsection{Architecture and training details}

The encoder is a \ac{ViT}-Base~\cite{ViT} with $d = 768$ dimensions while the predictor is a \ac{MLP} with $L = 6$ layers and $m = 1024$ hidden units.
To compute the CLAP embeddings, we use the pretrained \texttt{music} checkpoint from LAION.\footnote{\texttt{\url{https://huggingface.co/lukewys/laion_clap/blob/main/music_audioset_epoch_15_esc_90.14.pt}}}

We extract audio chunks of 8 seconds during training, which we convert into log-scaled Mel-spectrograms with 80 Mel bins, a window size of 25 ms and a hop size of 10 ms. We use $16 \times 16$ patches, thus a sequence size of $K = \frac{80}{16} \times \frac{800}{16} = 250$.

In both phases 1 and 2, we train our model during 300k steps using AdamW~\cite{AdamW}, with a batch size of 256, a base learning rate of 1e-3, and a cosine annealing scheduling after 20k steps of linear warmup.
All other hyperparameters are consistent with those used in \cite{StemJEPA}.

Each training phase takes approximately four days on a single A100 GPU with 40 GB of memory.

\subsection{Training data}

As in \cite{StemJEPA}, the model is trained on a proprietary dataset of 20,000 multi-track recordings of diverse music genres (rock, rap, R\&B, country...), totaling 1350 hours.
For most of the stems, we have access to a hierarchy of labels describing the instrument (e.g., ``guitar'' $\to$ ``electric guitar'' $\to$ ``lead electric guitar''). During phase 2, we pick the target label $c$ randomly from the hierarchy to make our model robust to both coarse and fine conditioning.
When no labels are available, we use the word ``music'' as conditioning.
Finally, we ensure not to pick silent chunks when sampling the context mix and target stem, as detailed in \cite{StemJEPA}.

\section{Experiments on musical stem retrieval}

We evaluate our model on a retrieval task using two publicly available datasets for source separation: MoisesDB~\cite{MoisesDB} and MUSDB18~\cite{MUSDB18}. The MoisesDB dataset contains 240 tracks divided into 38 \emph{instruments} across 11 \emph{categories} (drums, guitar, vocals...), totaling 2,585 stems. MUSDB18 consists of 150 tracks, each divided into exactly four stems: ``bass'', ``drums'', ``vocals'', and ``other''.

By evaluating our model on both datasets, we can assess how well the model handles conditioning at varying levels of granularity. Notably, MoisesDB includes instruments/categories that are absent from the training set, allowing us to test our model's zero-shot retrieval capabilities.

\subsection{Experimental setup}

We test the ability of our model to retrieve the missing stem from a given mix when conditioned on the instrument label of the missing stem.
Our dataset comprises $N$ tracks $\x^{(1)}, \dots, \x^{(N)}$, and each track $\xn$ is composed of $S_n$ stems $\xn_1, \dots, \xn_{S_n}$.

Given an individual stem $\xn_s$, $\xn_{\neg s}$ denotes the mix containing all stems from $\xn$ but $\xn_s$.
We first encode all individual stems and average the resulting embeddings over the patch dimensions to obtain a reference set $\mathbf{Z} = \{ \zn_s \}$ of latent representations.
%\TODO{Stefan: This sounds like you average over all individual stems which should result in one vector. Better "we average over the patch dimensions" or similar}.

In addition, we pass the mixes $\xn_{\neg s}$ through the encoder, whose outputs are fed into the predictor conditioned on the corresponding instrument label of $\xn_s$ to produce queries $\qn_s$. The queries are averaged similarly to the representations $\zn_s$. For such a query $\qn_s$, we then measure the distance to the representation of the corresponding stem $\zn_s$. To do so, we rely on two metrics:
%\TODO{present recall and normalized rank}
\begin{itemize}
    %\item The \emph{Recall at $k$} (R@$k$) measures the proportion of relevant items retrieved within the top $k$ nearest neighbors of the queries. We report the recall at $k$ of the different models for $k \in \{1, 5, 10\}$.
    \item The \emph{Recall at $k$} measures the proportions of queries $\qn_s$ for which the missing stem $\zn_s$ lies in their $k$ nearest neighbors. We report the recall at $k$ of the different models for $k \in \{1, 5, 10\}$.
    \item The \emph{Normalized Rank}~\cite{SampleMatch} of a query $\qn_s$ is the position of the ground-truth stem $\zn_s$ in the sorted list of distances $\{ \| \qn_s - \z \| \}_{\z \in \mathbf{Z}}$, normalized by the total number of stems $|\mathbf{Z}|$.
    For each model, we report the mean and median Normalized Ranks over all queries.
\end{itemize}

\subsection{Baselines}

There are no prior works specifically addressing stem retrieval conditioned on arbitrary instruments, so we lack a fair baseline for comparison. The closest related work is Stem-JEPA~\cite{StemJEPA}, but it is limited to four stems.

As a reference, we provide the performance of two models: one trained using only contrastive learning (equivalent to phase one of our approach) and the original Stem-JEPA model. For the contrastive baseline, since there is no conditioning, we use the (averaged) embeddings as queries directly.

To evaluate Stem-JEPA on MoisesDB, we map each instrument to one of the four standard sources. However, this process results in many instruments being grouped under the ``other'' category, which negatively impacts the performance on this dataset.

\subsection{Retrieval results}

We present our results in Table \ref{tab:retrieval}. For MUSDB18, the predictor is conditioned on text labels: ``bass'', ``drums'', ``vocals'', and ``music''. For MoisesDB, we report results both when conditioning the model on the actual instrument labels (fine conditioning) and on broader categories (coarse conditioning).

The results highlight the effectiveness of our design choices. 
On MoisesDB, using CLAP embeddings to condition the predictor on actual instrument/category labels significantly improves performance compared to Stem-JEPA: R@1 jumps from 9.9\% to 22.0\% or 19.2\%, respectively.
On MUSDB18, the performances are roughly similar (33.0\% vs. 33.5\% for R@1).
This can be attributed to the ``other'' category, where instrument specificity is less defined.

Interestingly, while contrastive learning performs poorly on R@1 (2.7\% for MoisesDB) compared to Stem-JEPA (9.9\%), it significantly outperforms it in all other metrics. This reveals that contrastive learning captures a solid global structure (with close projections for all stems of a track) but that the absence of conditioning avoids retrieving the right missing stem.
On all datasets, contrastive pretraining substantially boosts most metrics, but it also lowers R@1 (12.5\% vs. 19.3\% on MoisesDB with fine conditioning), indicating that the model's conditioning on the instrument label is being partially ignored by the predictor.

However, adding the FiLM conditioning at each layer of the predictor (rather than concatenating the embedding to its input) mitigates this effect.
Indeed, the latter yields the best overall performance, consistently outperforming other models, particularly for R@1 (22.0\% on MoisesDB) and median normalized rank (0.2\% on MoisesDB).
Finally, the close results on MoisesDB between the two conditioning levels show that using CLAP embeddings makes our model robust to finer and coarser conditioning.

\begin{figure*}
	{\hspace{0.1\textwidth}
		\includegraphics[width=0.8\textwidth]{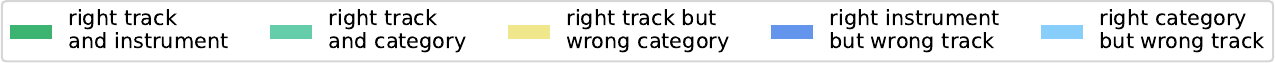}}
	
	\makebox[0pt][l]{\hspace{-1.6cm}
		\includegraphics[width=1.1\textwidth]{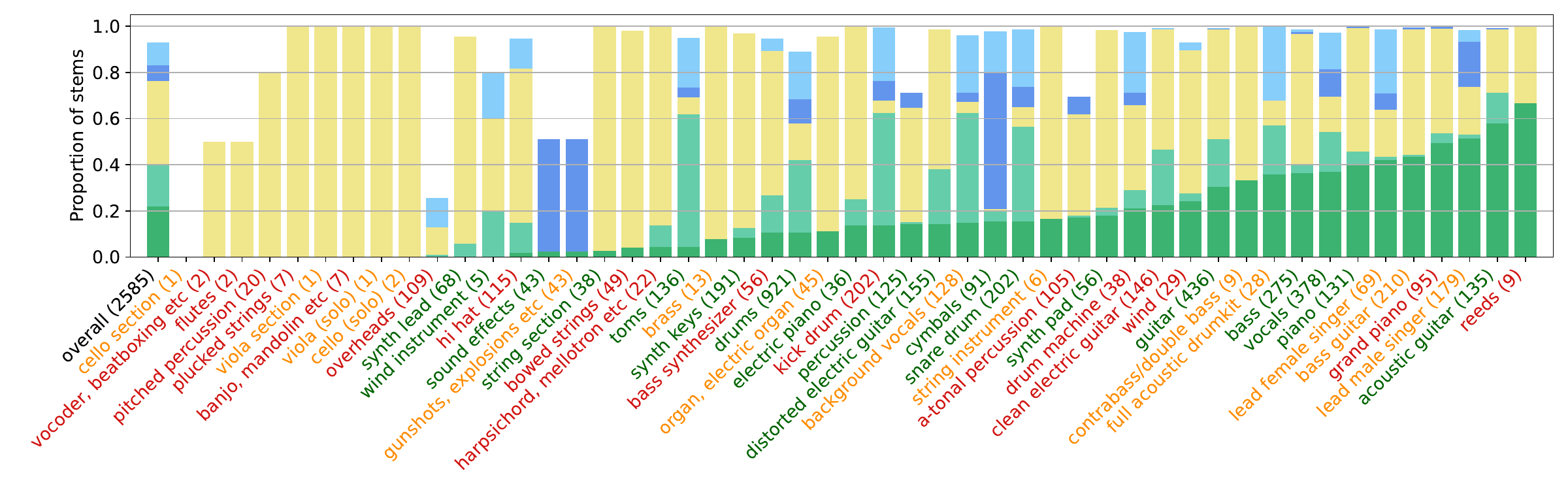}}
    % \captionsetup{font=\fontsize{9}{11}}
    \caption{%\normalsize{
        Analysis of the proportions of stems correctly retrieved by our model (with fine conditioning) on the MoisesDB dataset. The retrieved stem may come from the right audio track or not, and be the right instrument, another instrument from the same category (e.g. ``acoustic guitar'' vs. ``electric guitar'') or both wrong.
        The labels are the ones from MoisesDB and are colored as per their presence in the training set: green if they are in it, orange if they are but written differently (e.g., ``female lead vocals'' vs. ``lead female singer''), and red if they are not (zero-shot scenario).
    }
    \label{fig:moises}
\end{figure*}

\subsection{Instrument-specific analysis}

Metrics from Table \ref{tab:retrieval} assess how closely the latent representations of target stems align with their corresponding queries. In contrast, here we examine which stem is identified as the closest match to our queries.
We focus on analyzing the nearest-neighbor retrievals for queries on MoisesDB with fine conditioning. The analysis aims to explore both the types of failures our model encounters and the impact of the conditioning instrument on the retrieval performance.

Notably, some instruments are present in the training set, while others are either absent or labeled differently (e.g., ``lead female singer'' vs. ``female lead vocals''). Ideally, the structure of the CLAP embedding should allow the model to perform reasonably well even in zero-shot scenarios, which we evaluate in this analysis.

% Additionally, the taxonomy of our training set differs from MoisesDB. In MoisesDB, instruments that share the same label \gp{Geoffroy: je comprend pas la distinction entre label, caégories, ... il faut queue tu détails} (e.g., "vocoder," "beatboxing") may be categorized differently in our training data. Conversely, MoisesDB distinguishes between specific elements of the drum kit (e.g., kick, snare, hi-hat), while our model only sees a single stem for "drums" during training (\gp{Geoffroy: tu veux dire que dans voter base privée il n'y a que le label drum ?}. 
% It is, therefore, informative to assess how well our model performs for different instruments and to observe the types of failures that occur when it fails to retrieve the correct stem.

Our results are depicted on Figure \ref{fig:moises}. Overall, recall at 1 is only 22\%, but it jumps to 40\% when considering retrieving from the correct instrument category. Fully irrelevant results (wrong category and wrong track) are quite uncommon.
Performance is poor for underrepresented instruments (e.g., beatboxing, flutes, banjo). Interestingly, timbre has a larger effect than presence in the training set: the model performs well on piano, bass, vocals, and reeds (even though no stems are labeled "reeds"). However, despite seeing them during training, it struggles with wind instruments and sound effects.

The most common error is retrieving a stem from the same context mix instead of the target stem, likely influenced by contrastive pretraining. On the contrary, for sound effects, gunshots, and cymbals, the model often retrieves the right instrument but struggles to match them to the correct track, likely due to their similar timbral properties and the lack of harmonic cues.

When conditioned on specific drum kit parts (e.g., toms, kick, snare), the model frequently retrieves a different drum stem from the same track.
Indeed, in our training set there is usually a single stem for all drums, making it hard for the model to discriminate between the different drum stems.
A similar trend is observed for ``background vocals'' (labeled ``backing vocals'' in the training set). The model often retrieves a vocals stem from the correct track but struggles to distinguish the specific background vocal stem.

\section{Music representation learning}

In this section, we investigate the musical features encoded in the representations learned by our model. We hypothesize that the encoder captures shared musical information among different stems of the same track, such as rhythm or harmony, to aid the predictor. To verify this, we evaluate it on several downstream classification tasks, a standard protocol for representation learning methods~\cite{MULE,M2D,HEAR,MARBLE}.
%We train an MLP classifier on top of the representations returned by our frozen encoder.

Specifically, each audio sample from the downstream dataset is processed by the encoder, with its patch-wise outputs concatenated and averaged along the frequency and time dimensions to obtain a single 3840-dimensional global embedding per audio, as done in \cite{StemJEPA,M2D}.
%Each audio sample is processed by the \emph{frozen} encoder, and its patch-wise outputs are concatenated and averaged along frequency and time dimensions to produce a 3840-dimensional global embedding, following \cite{M2D}. 

These embeddings are passed through an MLP with 512 hidden units and a softmax layer, which is trained by minimizing the cross-entropy between the predicted distribution and the ground truth labels. To limit the bias induced by the optimization of the probe, results are computed as the average performance of three probes with identical hyperparameters but different seeds. In addition, we try different learning rates and report the best results, as described in the constrained evaluation track of \cite{MARBLE}.

\subsection{Downstream tasks} To validate our hypothesis, we focus on global musical features that are shared among the different stems of a track, namely tagging, key, and genre estimation.
Additionally, we include an instrument classification task to observe whether the encoder preserves stem-specific information. For facilitating comparisons to existing work, we also pick our downstream tasks from the MARBLE benchmark~\cite{MARBLE}. The full list of datasets and associated tasks is depicted in Table \ref{tab:tasks}.
For a more in-depth description of the datasets, tasks, and corresponding metrics, we refer the reader to \cite{MARBLE}. 

\begin{table}
	\centering
	\caption{Datasets used for downstream tasks.}
	\label{tab:tasks}
		\begin{tabular}{lcl}
			\toprule
			Dataset & classes & Task \\
			\midrule
			Giantsteps (GS) \cite{Giantsteps} & 24 & Key detection \\
			GTZAN \cite{GTZAN} & 10 & Genre classification \\
			MagnaTagATune (MTT) \cite{MTT} & 50 & Auto-tagging \\
			NSynth \cite{NSynth} & 11 & Instrument classification \\
			% & 88 & Pitch classification \\
			\bottomrule
		\end{tabular}
\end{table}

\subsection{Results}

Compared to the original Stem-JEPA restricted to 4 stems, we observe overall better performances.
Conditioning the predictor with CLAP embeddings consistently improves performances on all downstream tasks.

Contrastive pretraining, while effective for key-related tasks, leads to performance degradation on other tasks. For example, on GTZAN (genre classification) and MTT (auto-tagging), JEPA models with contrastive pretraining underperform compared to their standalone JEPA or contrastive baseline counterparts.
This suggests that our naive contrastive pretraining strategy—using identical hyperparameters and training durations for both contrastive and JEPA phases—may not be optimal. Adjustments such as fewer pretraining steps or higher contrastive loss temperatures could potentially yield better results.

Finally, conditioning with FiLM layers improves performances for all tasks.

\begin{table}
    \centering
    \caption{
        Influence of our different design choices on various downstream tasks, and comparison with existing baselines. Best results are bold and second-to-best are underlined.
    }
    \label{tab:downstream}
    {%\scriptsize
        \begin{tabular}{lccccc}
            \toprule
             & GS & GTZAN & \multicolumn{2}{c}{MTT} & NSynth \\
            Model & Acc\textsuperscript{refined} & Acc & ROC & AP & Acc \\
            \midrule
            Contrastive baseline &
            \textbf{66.4} & 68.9 & 89.4 & 41.6 & \textbf{76.8} \\
            %Contrastive no flat (ed948e48) &
            %63.5 & 65.3 & 67.9 & 80.7 & 88.5 & 38.6 & 75.1 \\            
            Stem-JEPA \cite{StemJEPA} &
            40.2 & 68.6 & 89.9 & \underline{42.8} & 73.5 \\
            \midrule
            +CLAP conditioning &
            50.0 & \underline{76.7} & \underline{90.5} & \textbf{44.0} & 75.5 \\
            %+FiLM conditioning (64969267) &
            %45.0 & \textbf{78.4} & 45.5 & 49.8 & 90.7 & \textbf{44.4} & 75.6 \\
            +contrastive pretraining &
            58.7 & 55.6 & 88.0 & 37.7 & 74.5 \\
            +FiLM conditioning &
            60.0 & 63.0 & 88.8 & 39.8 & \underline{76.7} \\
            % \textit{no flat} \\
            % +contrastive pretraining (990d42bb) &
            % 59.5 & 61.5 & 66.6 & 81.2 & 88.7 & 39.5 & 74.9 \\
            % +FiLM conditioning (0857dce1) &
            % 61.0 & 72.9 & 69.3 & 79.7 & 89.9 & 42.6 & 75.7 \\
            % +MCL (be32f074) &
            % 60.8 & 58.6 & 88.6 & 39.2 & 74.9 \\
            % \midrule
            % Stem-JEPA CLAP (72fe140f) &
            % 37.6 & 68.5 & 70.6 & 80.2 & 89.7 & 42.5 & 74.7 \\
            % Stem-JEPA w/o cond. (9cb7eb0d) &
            % 36.8 & 72.5 & 61.4 & 69.4 & 90.1 & 42.9 & 75.0 \\
            % Transformer (a12e52d8) &
            % 46.0 & 68.1 & 18.6 & 21.0 & 89.9 & 42.7 & 73.3 \\
            \midrule
            MULE \cite{MULE} &
            \underline{64.9} & 75.5 & 91.2 & 40.1 & 74.6 \\
            %MULE &
            %50.8 & 68.6 & 90.5 & 38.9 & 71.4 \\
            Jukebox \cite{Jukebox} &
            63.8 & \textbf{77.9} & \textbf{91.4} & 40.6 & 70.4 \\
            \bottomrule
        \end{tabular}
    }
\end{table}

\subsection{Comparison with baselines.}

Our models outperform MULE and Jukebox on most tasks, except auto-tagging on the MTT dataset. However, none of our models consistently surpass these baselines across all tasks.% In particular, models that excel on fine-grained tasks (key, tempo, pitch) often lag on high-level tasks (genre classification, tagging). This underscores a trade-off that suggests more careful optimization of pretraining strategies is necessary.

It is worth noting that both MULE and Jukebox were trained on significantly larger datasets—117k hours for MULE and 60k–120k hours for Jukebox—while our models were trained on just 1,350 hours of data, roughly 100 times less. This suggests that leveraging separated sources for music representation learning is a promising direction for reducing data requirements.

% Overall, our results highlight the difficulty of learning universal representations that are suited for all downstream tasks.
% Benchmarks like HEAR~\cite{HEAR} show that the best-performing models for general representation learning are often mixtures of several ones.
% Similarly, our Stem-JEPA models without contrastive pretraining excel in high-level tasks like genre classification and tagging but struggle with pitch, tempo, and key estimation.
% These tasks are well-defined and already addressed effectively by specialized methods.\footnote{In particular, the current SOTA methods for solving these three tasks with self-supervised learning are all inspired by our work, see \citet{PESTO,Gagnere2024,STONE}.}
% From a very practical standpoint, it is therefore uncertain whether it is even useful to try to solve all tasks at the same time.

\section{Temporal information in embeddings}

% \gp{il faut beaucoup réduire ce paragraph}
% % To evaluate the ability of our embeddings to capture local temporal information, we used them in a beat-tracking task. Each embedding represents a 160 ms audio segment, which lacks sufficient resolution to predict beat positions. To enhance temporal granularity, we employed a set of $K$ linear probes $ L_{k, \theta} $ ($0 \leq k < K$), where each head $k$ covers the sub-segment $\left[k \cdot \frac{0.160}{K}, (k+1) \cdot \frac{0.160}{K}\right]$ seconds. Setting K=8 achieved a 20 ms time resolution. Following standard methodology \cite{Bck2016JointBA, Bck2020DeconstructAR}, we conducted 8-fold cross-validation, reporting F-measure, AMLt, and CMLt metrics. Training and evaluation were performed on Ballroom, Hainsworth, SMC, and Harmonix datasets, with GTZAN reserved as an unseen test set. Probes were optimized using binary cross-entropy loss with label widening \cite{Bck2020DeconstructAR}. Results on GTZAN and SMC are shown in Table \ref{tab:beatTracking}.

Our previous results involve averaged embeddings and thus reveal the abilities of our model to capture \emph{global} musical information such as style, tempo, tonality...
In this part, we focus on whether \emph{local} information is also captured. To do so, we evaluate the sequences of embeddings learned by our encoder on a beat tracking task.
%This approach provides valuable insights, as beat tracking requires accurate detection of rhythmic patterns, allowing us to assess the embeddings’ ability to capture detailed temporal features.

\subsection{Experimental setup}

We concatenate the outputs of the encoder along the frequency axis. Each embedding encodes a \SI{160}{\milli\second} segment of the audio signal and is up-sampled using $K=8$ linear probes. Each probe targets a \SI{20}{\milli\second} sub-segment within the \SI{160}{\milli\second} window and outputs a beat activation function.
We use 8-fold cross-validation\footnote{We train on Ballroom \cite{ballroom}, Hainsworth \cite{hainsworth}, SMC \cite{SMC}, and Harmonix \cite{oriol_nieto_2019_3527870} and keep GTZAN \cite{GTZAN} reserved as an unseen test set.} and report F-measure, AMLt, and CMLt, following standard protocols \cite{Bck2016JointBA, Bck2020DeconstructAR}. The probes are trained with binary cross-entropy loss and label widening \cite{Bck2020DeconstructAR}.

To predict beat positions, the activation is fed into a dynamic Bayesian network \cite{florian_krebs_2018_1414966}.
% Results for GTZAN and SMC are presented in Table \ref{tab:beatTracking}. We compare our performances with a standard supervised method \cite{Bck2020DeconstructAR}, as well as another approach that also consists in \ac{SSL} pre-training followed by linear probing~\cite{contrastiveBeatTracking}.
% %, and that follows the same evaluation protocol.

% \input{tables/beat-tracking}

% Our model achieves decent results, with AMLt scores close to those of the baselines.
% Even if not as good as the other methods, especially in terms of F-measure and CMLt, these results reveal that, despite a patch resolution of only 160 ms, the embeddings learned by our model retain precise temporal information. This makes our approach promising for more fine-grained tasks.
 
We compare our method against two baseline approaches:
a supervised beat tracking model~\cite{Bck2020DeconstructAR}, optimized specifically for this task
and a self-supervised approach using contrastive pretraining followed by linear probing~\cite{contrastiveBeatTracking}, evaluated under the same protocol as our method.

%The baseline models represent strong benchmarks for assessing the performance of our approach.

\subsection{Results}

\begin{table}%[ht]
    \centering
    \caption{
    Beat-tracking performance in 8-fold cross-validation setting.
    }
    \label{tab:beatTracking}
    \setlength{\tabcolsep}{3pt} % Adjust column separation
    %\scriptsize % Reduce font size
    {%\footnotesize
        \begin{tabular}{lcccccccc}
            \toprule
             &  & \multicolumn{3}{c}{GTZAN} &  & \multicolumn{3}{c}{SMC} \\
            \cmidrule{3-5} \cmidrule{7-9}
            Model &  & F1 & AMLt & CMLt &  & F1 & AMLt & CMLt \\
            \midrule
            % \cite{StemJEPA} + CLAP conditioning & .850 & .921 & .753 & .536 & .643 & .427\\
            Ours &  &
            0.845 & 0.917 & 0.741 &   & 0.534 & 0.654 & 0.440 \\
            Gagneré et al. \cite{contrastiveBeatTracking} &  &
            0.876 & 0.918 & 0.802 &  & \textbf{0.567} & 0.651 & 0.480   \\ 
            \midrule
            Böck et al. \cite{Bck2020DeconstructAR} &  &
            \textbf{0.885} & \textbf{0.931} & \textbf{0.885} &  & 0.552 & \textbf{0.663} & \textbf{0.514} \\
            \bottomrule
        \end{tabular}
    }
\end{table}

Table~\ref{tab:beatTracking} presents the results on GTZAN and SMC datasets. Our model achieved reasonable results, particularly in terms of AMLt, where scores were close to those of the baselines. However, the model underperformed compared to the baselines in terms of F-measure and CMLt.  

Despite its limitations, these results demonstrate that, even with a patch resolution of \SI{160}{\milli\second}, the embeddings learned by our self-supervised model retain detailed temporal information. This suggests that our approach is promising for tasks requiring finer temporal precision, highlighting its potential for further refinement and application.

% \input{tables/beat-tracking}

% \begin{table*}
%     \centering
%     \caption{
%     Beat detection F1-score computed across each dataset
%     }
%     \label{tab:beatDetection}
%     {\footnotesize
%         \begin{tabular}{lcccccc}
%             \toprule
%              Model & Ballroom & GTZAN & Hainsworth & Harmonix & RWC & SMC \\
%             \midrule
%             6a70 & 0.842 & 0.801 & 0.851 & 0.858 & 0.688 & 0.547
%         \end{tabular}
%     }
% \end{table*}

\section{Conclusion}

In this paper, we introduce the first model for zero-shot musical stem retrieval using \ac{JEPA}s. We build upon the work of \cite{StemJEPA} and extend it to arbitrary instruments, improving its usability in real-world scenarios. Additionally, we explore alternative design choices for the predictor, replacing concatenation with FiLM conditioning, and propose pre-training the encoder via contrastive learning.

Our results indicate that these modifications not only generalize Stem-JEPA but also enhance its performance, both on retrieval tasks even in the 4-stem configuration (see Table~\ref{tab:retrieval}), and across various MIR downstream tasks, occasionally outperforming top-performing baselines (see Table~\ref{tab:downstream}).

However, our experiments reveal the challenge of consistently achieving the best of both worlds. Across multiple tasks, there is no single design choice that emerges as clearly superior. This suggests that further research is necessary to identify an optimal balance, both in terms of pre-training strategies and predictor conditioning methods.

%Finally, while our work represents a significant step forward, it is not without limitations. We outline these limitations below and suggest directions for future research to address them.

%In this paper, we introduce the first model for zero-shot musical stem retrieval using \ac{JEPA}s. We demonstrate the effectiveness of our design choices by evaluating our model on different benchmarks, and we significantly outperform previous baselines.
%Interestingly, 

In addition, as our results on beat tracking reveal, our embeddings are not only useful for musical stem retrieval but also retain temporal information, making them potentially usable for both global and local MIR tasks or music accompaniment generation.

More generally, despite our study mostly focuses on stem retrieval, pretraining a \ac{JEPA} with contrastive learning and/or using a pretrained multimodal model for conditioning its predictor is not limited to the music domain. Our design choices may, therefore, be useful to improve performances on other tasks beyond the scope of this paper.
% \begin{itemize}
%     \item First model for zero-shot musical stem retrieval
%     \item Good performances on very challenging benchmarks and outperforms previous systems on simpler ones
%     \item In addition, we show that the learnt embeddings retain local information, making them potentially usable for other tasks such as stem alignment or conditional music generation.
%     \item Finally, we are the firsts to combine contrastive learning and JEPA, potentially other applications beyond musical stem retrieval, such as representation learning.

% \end{itemize}

%\newpage

\section{Acknowledgments}

This work has been funded by the ANRT CIFRE convention n°2021/1537 and Sony France. This work was granted access to the HPC/AI resources of IDRIS under the allocation 2022-AD011013842 made by GENCI.

We would like to thank Amaury Delort for his support regarding data management and preprocessing.

\bibliographystyle{IEEEbib}
\bibliography{strings,refs}

\end{document}